\let\latexaddtocontents\addtocontents
\documentclass[submitting, floatfix]{nst}
\let\addtocontents\latexaddtocontents
\usepackage{natbib}
\usepackage{subfigure,dcolumn}
\usepackage[T2A,T1]{fontenc}
\usepackage[english]{babel}
\usepackage{color}
\usepackage{listings}
\usepackage{textcomp,mathcomp}
\usepackage{comment}
\usepackage{lineno}

\let\footnote\relax

\let\textcite\relax

\let\citeauthor\relax
\let\citeyear\relax
\expandafter\let\csname
ver@natbib.sty\endcsname\relax

\let\linenumbers\relax
\usepackage[printonlyused,nohyperlinks]{acronym} 

\usepackage[
    backend=biber,
    bibencoding=utf8,
    style=numeric-comp,
    giveninits=true,
    autolang=other, 
    sorting=none,
    doi=true, isbn=false, eprint=false, url=false, date=year, 
    maxcitenames=3,
    mincitenames=3,
    maxbibnames=3,
    minbibnames=3,
]{biblatex}
\usepackage{csquotes}
\renewbibmacro{in:}{}
\begin{document}
\renewcommand{\bibliography}[1]{}

\title{Unity based virtual reality for detector and event visualization in JUNO experiment}

\author{Kai-Xuan Huang}
\thanks{These authors contributed equally to this work.}
\affiliation{School of Physics, Sun Yat-sen University, Guangzhou 510275, China}
\author{Tian-Zi Song}
\thanks{These authors contributed equally to this work.}
\affiliation{School of Physics, Sun Yat-sen University, Guangzhou 510275, China}
\author{Yu-Ning Su}
\affiliation{School of Physics, Sun Yat-sen University, Guangzhou 510275, China}
\author{Cheng-Xin Wu}
\affiliation{School of Physics, Sun Yat-sen University, Guangzhou 510275, China}
\author{Xue-Sen Wang}
\affiliation{Sino-French Institute of Nuclear Engineering and Technology, Sun Yat-sen University, Zhuhai 519082, China}
\author{Yu-Mei Zhang}
\email[Yu-Mei Zhang, ]{zhangym26@mail.sysu.edu.cn}
\affiliation{Sino-French Institute of Nuclear Engineering and Technology, Sun Yat-sen University, Zhuhai 519082, China}
\author{Zheng-Yun You}
\email[Zheng-Yun You, ]{youzhy5@mail.sysu.edu.cn}
\affiliation{School of Physics, Sun Yat-sen University, Guangzhou 510275, China}

\begin{abstract}
Detector and event visualization are crucial components of high-energy physics~(HEP) experimental software. 
Virtual Reality~(VR) technologies and multimedia development platforms such as Unity offer enhanced display effects and flexible extensibility for visualization in HEP experiments. 
In this study, we present a VR-based method for detector and event displays in the Jiangmen Underground Neutrino Observatory~(JUNO) experiment. 
This method shares the same detector geometry descriptions and event data model as those in offline software and provides necessary data conversion interfaces.
The VR methodology facilitates an immersive exploration of the virtual environment in JUNO, enabling users to investigate detector geometry, visualize event data, and tune the detector simulation and event reconstruction algorithms. Additionally, this approach supports applications in data monitoring, physics data analysis, and public outreach initiatives. 

\end{abstract}

\maketitle

\section{Introduction}
Visualization techniques are essential in every aspect of modern \ac{HEP} experiments. 
In the Roadmap for HEP Software and Computing R\&D for the 2020s~\cite{Roadmap} and the HEP Software Foundation Community White Paper~\cite{White_Paper_Visualization}, recommendations and guidelines for visualization tools, such as \ac{VR} technologies~\cite{VR1} in future software development are specifically discussed, particularly regarding interactivity, detector geometry visualization and event display.
Compared to traditional visualizations, VR techniques offer a truly immersive perspective, which enhances interactive experience with a better understanding of detector geometry and event information.
In recent years, some HEP experiments have developed VR applications for event display and outreach.
These include the Belle2VR software~\cite{Belle2_VR1, Belle2_VR2} for the BelleII experiment~\cite{Belle2}, the ATLASrift platform~\cite{ATLAS_VR1, ATLAS_VR2} for the ATLAS experiment~\cite{ATLAS}, the CMS VR application~\cite{CMS_VR} for the CMS experiment~\cite{CMS}, and the Super-KAVE program~\cite{SuperK_VR1, SuperK_VR2} for the Super-K experiment~\cite{SuperK}.

The development of VR applications typically involves game engines, such as Unity~\cite{Unity} or Unreal Engine~\cite{Unreal}.
Unity is a cross-platform engine that supports the development of games, videos, animations, and architectural visualizations.
It has been employed for detector visualization and event display in various HEP experiments, including BelleII, BESIII~\cite{BESIII_unity}, ALICE~\cite{ALICE}, ATLAS~\cite{CAMELIA_web}, JUNO~\cite{ELAINA}, and \ac{TEV} of the CERN Media Lab~\cite{CERNMediaLab}, all of which achieve excellent visualization effects.

The \ac{JUNO}~\cite{JUNO1,JUNO2,juno_yellow_book} is situated underground in southern China with a 650~meter rock overburden.
The primary scientific goal of JUNO is to determine the neutrino mass hierarchy.
Over an approximately seven–year operational period, JUNO is expected to determine the neutrino mass hierarchy with a significance of $3\sigma$~\cite{NMO}, and to measure the oscillation parameters $\Delta m^2_{31}$, $\Delta m^2_{21}$, and $\sin^2\theta_{12}$, achieving a precision of $0.2\%$ for $\Delta m^2_{31}$, $0.3\%$ for $\Delta m^2_{21}$, and $0.5\%$ for $\sin^2\theta_{12}$~\cite{JUNO_accuracy, Joao_Proceedings}, respectively.

Additionally, JUNO experiment is capable of investigating various types of neutrinos, including earth neutrinos, atmospheric neutrinos, solar neutrinos and supernova neutrinos~\cite{JUNO1}.
Its excellent energy resolution and large fiducial volume provide promising opportunities to explore numerous essential topics in neutrino physics.

In this study, we develop a VR-based event display tool using Unity for JUNO.
This software is compatible with various platforms through \ac{HMDs}~\cite{HMDs} and offers functionalities including the VR-based visualization of JUNO detector, event displays for different types of data, interfaces for reading and converting event data information, and spatial \ac{Spatial UI} control features.

The rest of this paper is organized as follows.
In Section~\ref{sec: Visualization and VR}, we introduce VR-based software for HEP experiments. 
In Section~\ref{sec: methodologies}, the software methodologies is described, including the JUNO VR framework, the data flow of detector geometry and event data conversion, as well as interaction methods with Spatial UI. 
The visualization of detector units and event data in the VR-based tool is introduced in Section~\ref{sec: Visualization of detector and events}.
The potential for further applications is discussed in Section~\ref{sec: Applications}.
Finally, the performance of the software is introduced in Section~\ref{sec: Performance}.

\section{Visualization and VR}
\label{sec: Visualization and VR}
\subsection{ Unity and VR}
In HEP experiments, physicists typically develop detector descriptions and event visualization tools within offline software frameworks.
These event display tools are usually built upon the widely-used HEP software, such as Geant4~\cite{GEANT4} or ROOT~\cite{ROOT}, which provide user-friendly visualization capabilities that facilitate software development. 
With the upgrades to ROOT and its EVE package~\cite{ROOT_EVE}, the development of event display tools has become more efficient.
Several recent HEP experiments, including ALICE, CMS~\cite{CMS}, BESIII~\cite{zhijun_Visualization_BESIII_ROOT}, JUNO~\cite{JUNO_event_display_ROOT, TAO_ROOT}, and Mu2e~\cite{Mu2e}, adopt ROOT EVE for developing event display software.
However, due to the limited visualization technique support in ROOT, its display capabilities do not fully meet the diverse requirements of physicists, and most ROOT applications remain confined to the Linux platform.

To enhance visualization quality, interactivity, and multi-platform support, several event display tools are developed based on external visualization software.
Unity is widely applied in the field of HEP, being used in projects including BelleII, BESIII, ALICE, ATLAS, and JUNO.
Unity is a professional video and game development engine based on C\#, and visualization software built on Unity offers several advantages.

\begin{itemize}
\item \emph{Impressive visualization quality.}
Unity, a widely adopted professional 3D engine in the industry, offers advanced visual capabilities that surpass those of traditional software used in HEP, such as ROOT.
Additionally, its continuous updates enable HEP visualizations to stay aligned with the cutting-edge developments in graphics technology.

\item \emph{Cross-platform support.} 
The comprehensive multi-platform support of Unity enables seamless export and deployment of projections across a range of operating systems, including Windows, Linux, macOS, iOS, Android, and web browsers. 
This functionality ensures that the same visualization project can be accessed across various platforms, minimizing the development effort and streamlining maintenance tasks.

\item \emph{High-quality VR rendering and performance optimization.} 
Unity supports modern graphics technologies such as real-time lighting, global illumination, and physics-based rendering. 
Light behaves according to the principles of physics, including energy conservation and Fresnel reflections~\cite{unity_standard_shader}, resulting in more realistic and immersive graphical effects in VR.
These features are crucial for enhancing details like lighting, shadows, textures, and environmental interactions, significantly improving the user’s sense of immersion.
Additionally, Unity optimizes VR performance by rendering separate images for each eye, providing a dual-eye perspective while maintaining smooth rendering and minimizing motion blur and latency.

\item \emph{VR HMDs compatibility.} 
Unity supports most popular VR HMDs, including Meta Quest~2 and Quest~3~\cite{quest3}, HTC Vive~\cite{htcvive}, Valve Index~\cite{valveindex}, and Vision Pro~\cite{VisionPro}.
Through the extended reality interaction toolkit in Unity, developers can easily create interactive applications for various devices without device-specific coding.
\end{itemize}

Additionally, Unity provides a fast turnaround during the development cycle.
Projects can be executed immediately, running quickly on VR devices for easier debugging, without the need to compile and link executable files~\cite{unity_manual}.

Compared to 3D-based event visualization software, VR technology significantly enhances the visual experience of the user. 
VR applications are typically conducted through HMDs.
According to Steam VR hardware statistics~\cite{Steam_Hardware}, more than half of users utilize the Meta Quest~2 and Quest~3. 
These devices, based on the Android operating system, offer sufficient immersion and are widely used across various fields, including gaming, social interaction, and education.
Equipped with accelerometers, gyroscopes, and cameras, these devices can track the user's head and hand movements, enabling interaction and navigation within virtual environments.
Additionally, the controllers facilitate interaction with Spatial UI in the virtual environment.
VR technology provides synthesized sensory feedback, creating a strong sense of immersion and presence within a simulated environment.

Most HEP experiments are typically conducted in underground or restricted areas, which are typically inaccessible during data collection.
VR technology enables the public to explore these experiments in an immersive environment to observe detector operations and event data collection.
This offers a fundamental understanding of the types of scientific research being conducted in HEP, which is highly beneficial for both educational and outreach purposes.

Furthermore, by simulating particle emissions and their interactions with detectors, VR provides physicists with an immersive platform for refining offline simulations and reconstruction software~\cite{rec1, rec2, rec3, rec4}.
It can also enhance simulation accuracy.
For JUNO, considering the deformation of the stainless steel truss, offsets need to be applied to the PMT positions based on limited survey data~\cite{CD, juno_sim, Dengziyan_JUNO}.
Overlap checks and position tuning using the VR event display tool will be particularly helpful. 
Additionally, VR enables physicists to analyze rare events as though they are physically present within the inner detector environment, which provides alternative approach for data analysis and may inspire creativity.

\subsection{ VR application in HEP}
In recent years, VR applications have been developed in several HEP experiments event visualization and outreach.
These software include Belle2VR~\cite{Belle2_VR2} for the BelleII experiment, ATLASrift~\cite{ATLAS_VR1, ATLAS_VR2} for the ATLAS experiment, and Super-KAVE~\cite{SuperK_VR1, SuperK_VR2} for the Super-K experiment.



Belle2VR is an interactive VR visualization tool developed with Unity, designed to represent subatomic particle physics.
This application allows user to explore the BelleII detector and observe particle jets generated in high energy $e^+e^-$ collisions.
The Super-KAVE application immerses user in a scaled representation of the Super-K detector, allowing them to explore the virtual space, switch between event datasets, and change visualization modes~\cite{SuperK_VR1, SuperK_VR2}.
In addition to providing VR modes for exploring the detector and standard event displays, the application features a supernova event visualization technique, simulating the conversion of a star into a supernova.
This leads to the occurrence of thousands of neutrino events within approximately ten seconds.
It serves as a valuable outreach tool, offering a new example of visualization techniques for various neutrino particle physics applications.
ATLASrift, a VR application developed for the ATLAS experiment, is primarily used for data visualization and outreach~\cite{ATLAS}. 
User can move around and inside the detector, as well as explore the entire underground experimental cavern and its associated facilities, including shafts, service halls, passageways, scaffolds, and more.

\section{Methodologies}
\label{sec: methodologies}
VR technology provides an immersive experience. 
However, the development of comprehensive event display software utilizing VR for HEP experiments still involves significant challenges.

The first challenge is to convert the detector geometry, typically based on Geant4 simulations, into a format such as FBX~\cite{FBX} that can be imported into Unity.
Given that detectors usually consist of tens of thousands of components, manually creating the geometry would impose a significant workload.
Another significant challenge is extracting and converting event information into a structure compatible with Unity.
In HEP experiments, the fundamental information for event display is typically defined by the offline software and stored in ROOT format.
However, because Unity does not support direct reading of ROOT files, a dedicated conversion process is required.
Additionally, a bijective mapping is established to link the detector unit identifiers used in the offline software~\cite{Identifier_Moudle} with the names assigned to the corresponding geometries in Unity.

This section introduces the software architecture and data flow in the JUNO VR program.
We describe the process of detector geometry conversion, the exchange of essential event information from offline software to Unity, and the strategy of matching detector units.
Additionally, we discuss the construction of the Spatial UI and provide an overview of its functionality.

\begin{figure*}[!htb]
	\includegraphics[width=0.75\hsize]{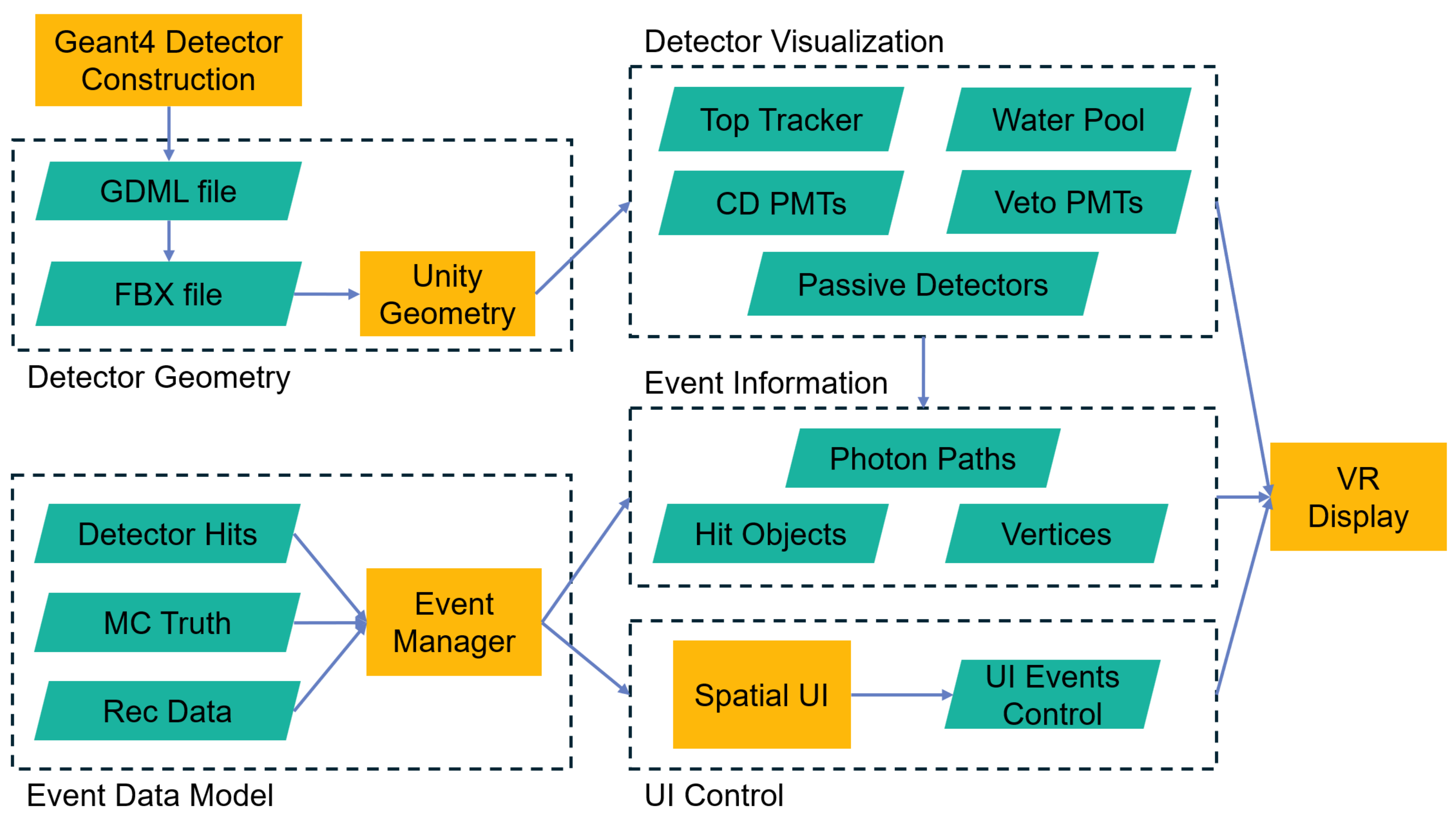}
	\caption{The software framework and data flow in JUNO VR.}
	\label{fig:JUNO_VR_Framework}
\end{figure*}

\subsection{Software structure and data flow}
The event display software should provide visualization capabilities, including detector geometry, event data information at different levels, and interactive controls.
For JUNO VR software visualization, the first step involves converting and importing the detector geometry and event data information into Unity for display, followed by the development of interactive controls.
As shown in Figure~\ref{fig:JUNO_VR_Framework}, the JUNO event display software consists of four components.

\begin{itemize}
\item \emph{Detector geometry conversion.}
The geometric models of the detector are constructed using Geant4 in the detector simulation, initially stored in a \ac{GDML} file~\cite{GDML}.
The GDML file is then automatically converted to the FBX format using the GDML-FBX conversion tool~\cite{BM@N, BESIII_unity}, which is compatible for import into Unity.

\item \emph{Event data conversion.}
The \ac{EDM} \cite{EDM} encompasses various types of event information exchanged between different components of JUNO online and offline software, including data acquisition, simulation, calibration, and reconstruction.
The event information for JUNO VR event display is extracted from the offline software EDM~\cite{JUNO_offline_software}.
By combining the detector identifier and Unity geometry name matching rules, the detector information is remapped, generating event information that Unity can directly import and conforms to the geometry hierarchy in Unity.

\item \emph{Detector and event information visualization.}
The detector geometry, simulation, and reconstruction information, as well as the hit information and their associations, are visualized in Unity.
By adjusting the material properties and combining Unity's layers, lighting, and rendering effects, an immersive and outstanding visualization experience in VR mode is achieved.

\item \emph{Spatial UI and interactive control.}
The Spatial UI is designed to facilitate the visualization and interaction with the detector and event information.
It includes the sub-detector geometry panel and the event display panel, which allow users to control the display of sub-detectors, switch between event types, and manage the event display process.
Interactive control is enabled through the Meta Quest~3 controller, with distinct functions assigned to the joystick and various buttons.
These functions include controlling the visibility of each panel, navigating within the 3D virtual detector environment, and switching perspectives.

\end{itemize}

\subsection{Detector geometry conversion}
\label{subsec: Detector Geometry Conversion}
The detector geometry in HEP experiments are typically complex, consisting of up to millions of detector units.
The description of these detectors is commonly developed using specialized geometric languages, such as GDML and \ac{DD4hep}~\cite{DD4HEP1,DD4HEP2}.
The JUNO experiment, along with BESIII, PHENIX \cite{PHENIX}, and LHCb \cite{LHCb}, uses GDML to describe and optimize the geometry of detectors for conceptual design and offline software development.
GDML is a detector description language based on \ac{XML}~\cite{XML} that describes detector information through a set of textual tags and attributes, providing a persistent description of the detector.
The geometry description files of detectors typically include essential information about the detector model, such as lists of materials, positions, rotations, solids, and the hierarchical structure of the detector.

Since the GDML format does not directly support import into Unity, some previous HEP applications involving Unity typically required manual construction of geometric models.
Given that HEP detectors are usually highly complex, the creation of 3D detector models in Unity becomes particularly challenging.
However, Unity does support direct import of several 3D file formats, including FBX, DAE~\cite{DAE}, DXF~\cite{DXF}, and OBJ~\cite{OBJ}.
Among these, FBX stands out as a widely used 3D asset format, due to its ability to handle intricate scene structures. 
This includes not only geometry but also animations, materials, textures, lighting, and cameras, which makes it a highly suitable choice for HEP applications involving complex 3D models.

A method that can automatically convert GDML or DD4hep to FBX format is essential for detector construction in Unity.
Several researches have proposed automated methods for converting GDML files to FBX files, significantly facilitating Unity-based development.
For instance, the BESIII collaboration group suggests using FreeCAD~\cite{FreeCAD}, a 3D CAD and modeling software, in conjunction with CAD data optimization software Pixyz~\cite{pixyz}, with the STEP~\cite{STEP} format as an intermediate conversion format~\cite{BESIII_unity}.
The CMS collaboration group employ SketchUp software for auxiliary data conversion~\cite{CMS_SketchUp}.

Recently, methods were also proposed to directly convert GDML files to FBX files~\cite{BM@N}.
This research, based on the above method, enables a fast and automatic conversion process from GDML to FBX, which can be completed in just a few minutes, and saves significant time in the conversion process.
This approach becomes particularly beneficial during the recent geometric updates of JUNO detector at the commissioning stage, enabling the swift conversion of the updated FBX file, which includes the latest geometry model of the real detector units after installation.

\subsection{Event data conversion }
In HEP experiments, the event data is typically stored in files with binary raw data format or ROOT format.
ROOT, an efficient data analysis framework, is widely adopted for high-performance data input and output operations. 
However, since Unity cannot directly read ROOT files, it is necessary to extract the required event information based on the EDM and convert it into a text format that Unity can process.

The essential information for event display comprises three main components: detector unit hits, \ac{MC} truth, and reconstruction data.
The detector unit hits include the hit time and hit charge for each detector unit like a PMT.
MC truth provides detailed truth information such as simulated vertices and photon trajectories (including 3D coordinates and propagation with time), which facilitate a deeper analysis of particle direction and relative velocity.
Reconstruction data typically contain the reconstructed vertex positions, energy information, and additional track information for muon events like direction.
Together, these information serve as the foundation for developing event display functionalities and interactive control modules based on Spatial UI.

Furthermore, the identifiers used for detector units in offline software may differ from the names of the geometric objects in Unity. 
In HEP experiments, the detector identifier system assigns a unique ID to each detector unit and play a critical role in various applications including data acquisition, simulation, reconstruction and analysis.
Therefore, establishing an accurate mapping between the detector identifiers in offline software and the geometric objects like PMT in Unity is essential to ensure the accurate display of an event.
Based on EDM readout rules and leveraging the mapping between the identifier module and the geometric objects in Unity, an automated readout and conversion interface is developed to export event display information.

For JUNO VR software, multiple types of datasets are provided, including radioactive background, \ac{IBD}~\cite{IBD}, cosmic ray muons and other types of events.
The event display dataset is designed to encompass both simulated and real data event types.
Simulated events are produced with the JUNO offline software to facilitate detector simulation, commissioning and the optimization of reconstruction algorithms. Since JUNO has not yet commenced formal data taking, real data events are instead obtained from the Data Challenge dataset~\cite{DC1}, which has data structures identical to those expected during actual operation.
With the event data conversion interface, the datasets with various types of data are ready to be displayed in the Unity based visualization and VR software.

\begin{figure}[!htb]
	\includegraphics[width=0.9\hsize]{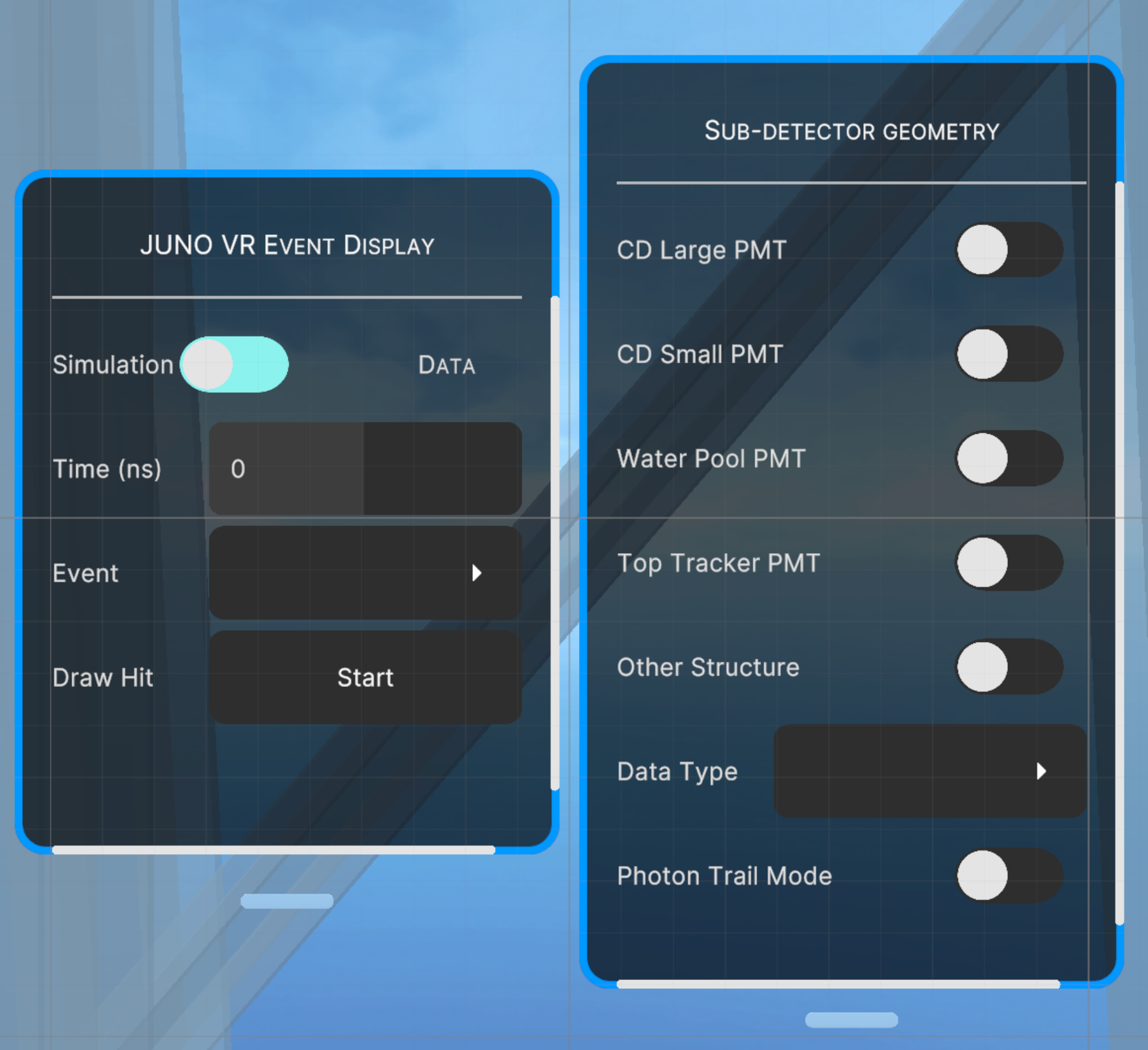}
        \caption{The Spatial UI in JUNO VR. On the left is the JUNO VR event display control panel, and on the right is the sub-detector geometry control panel.}
	\label{fig:Spatial UI}
\end{figure}

\subsection{Spatial UI and interactive control}
The Spatial UI serves as the interface facilitating interaction between user and the VR application.
For JUNO VR project, we develop two Spatial UIs: the sub-detector geometry control panel and the event display control panel, as shown in Figure~\ref{fig:Spatial UI}.

The sub-detector geometry panel primarily controls the visualization attributes of the geometries of various sub-detectors, including \ac{CD} large PMTs, CD small PMTs, Top Tracker, and water pool PMTs.
Detailed information about the sub-detectors of JUNO is provided in Section~\ref{subsec:Detector units}.
In addition to the sensitive detectors like PMTs, an "Other structure" toggle controls the display of passive structures, such as the steel structure, the acrylic ball, the PMT support structures, and the liquid filling pipelines.
Additionally, the "Data type" drop-down is used to switch between different types of events collected during real data-taking or from simulation.
The "Photon trail mode" toggle enables the switching of display modes for photon paths, either represented by green lines or in a manner closely resembling particle motion.

The event display panel is designed to implement the core functionality for event visualization, which includes a toggle for switching display mode between simulation and data types, a slider for controlling the display of an event with its timeline evolution, a drop-down for selecting different types of events, and a button to play the event animation.
A "Draw Hit" button initiates the animation of the full event hit process, which plays within a period of time window, with the time slider moving in sync with the event timeline, enabling user to track the current time of the event.

Interactive control is achieved through the use of controllers, gesture operations, eye-tracking, and other input methods in HMDs.
The following discussion focuses on testing interactive control for the Meta Quest~3. 
For other HMDs, the cross-platform support provided by the extended reality interaction toolkit in Unity minimizes development differences between various devices.
Simple adaptations based on the specific features of the HMDs are sufficient for operation.

The controller buttons resemble a typical gamepad, with the addition of side buttons.
The X\&Y buttons on the left controller are used to control the visibility of the sub-detector geometry panel.
When displayed, the position of this panel is based on the user's orientation and appears at the front left of the user’s view. 
User can drag or hide the panel to avoid obstructing their view when visualizing events.
The A\&B buttons on the right controller are used to control the visibility of the event display panel.
When displayed, the panel appears at the front right of the user's view.
Based on the gyroscope and accelerometer hardware of the Meta Quest~3, these planes are always oriented perpendicular to the user's view orientation.

\begin{figure}[!htb]
	\includegraphics[width=0.9\hsize]{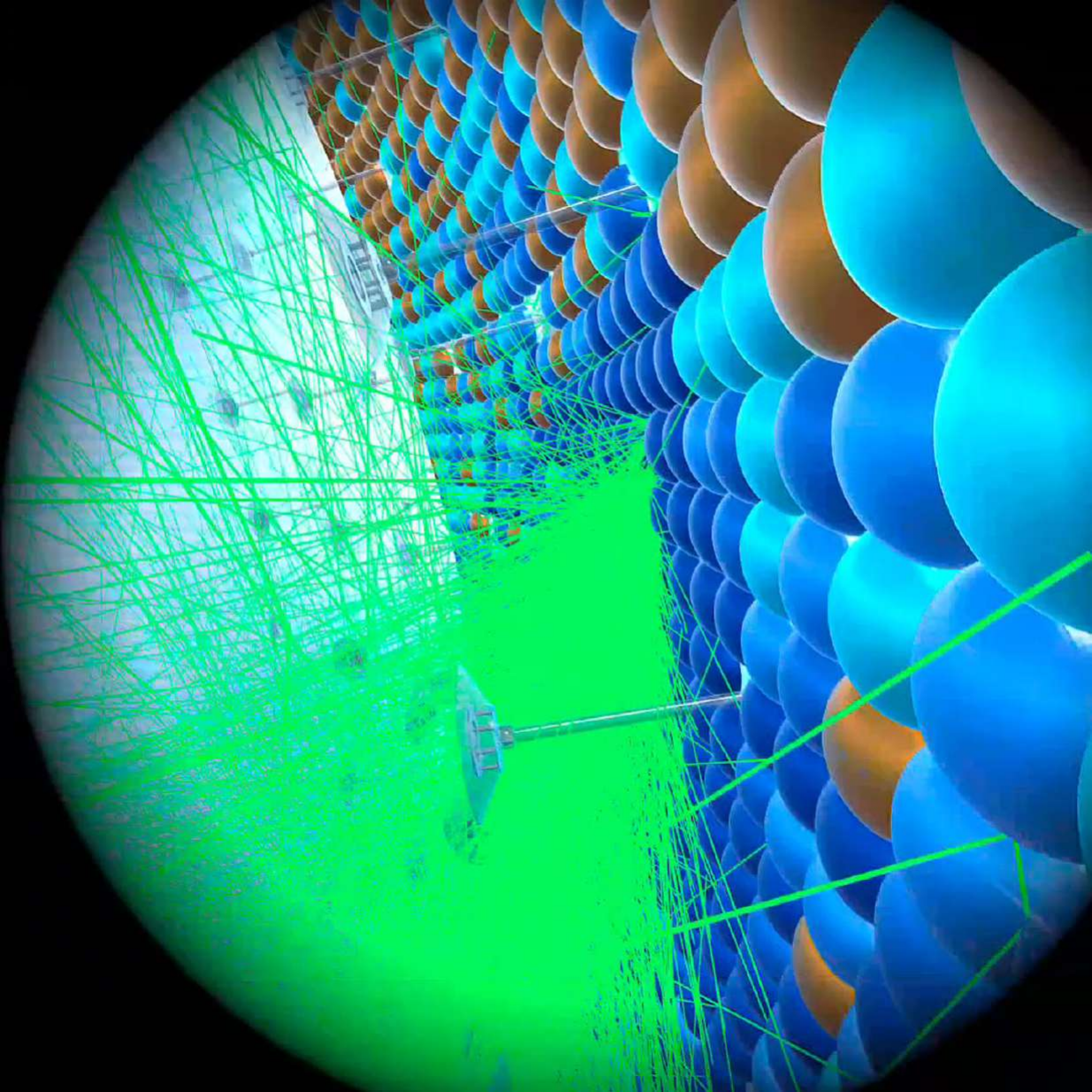}
	\caption{
    User perspective during motion while checking the display information of a simulated muon event in the JUNO VR application.
    The CD small PMTs are not shown.
    Detailed information about the sub-detectors of JUNO is provided in Section~\ref{subsec:Detector units}.
    }
	\label{Move_in_VR}
\end{figure}

The joystick on the left controller controls the user’s 3D movement, based on both the controller's input and the user's view orientation. 
For example, when the user’s head orientation is directed towards the upper right, pushing the joystick upwards move the user in the virtual space toward that direction.
Figure~\ref{Move_in_VR} illustrates the user’s viewpoint during motion in the JUNO VR application.
The event depicted is a simulated muon event. 
Additional details presented in the figure are described in detail in Section~\ref{sec: Visualization of detector and events}.
The joystick on the right controls the user’s viewpoint direction.
Additionally, the user can change their head orientation to switch perspectives.
The side button is used for interaction confirmation. 
Furthermore, when interacting with the Spatial UI, if the controller's laser pointer touches the corresponding component, audio and highlight feedback are provided, making the interaction smoother for user control.

\section{Visualization in JUNO}
\label{sec: Visualization of detector and events}
This section is dedicated to introduce the visualization effects in the JUNO VR, including detector geometry, hit distribution for different types of events, as well as  MC true information and display of event reconstruction outputs.

\subsection{Detector units}
\label{subsec:Detector units}

\begin{figure}[!htb]
	\includegraphics[width=0.99\hsize]{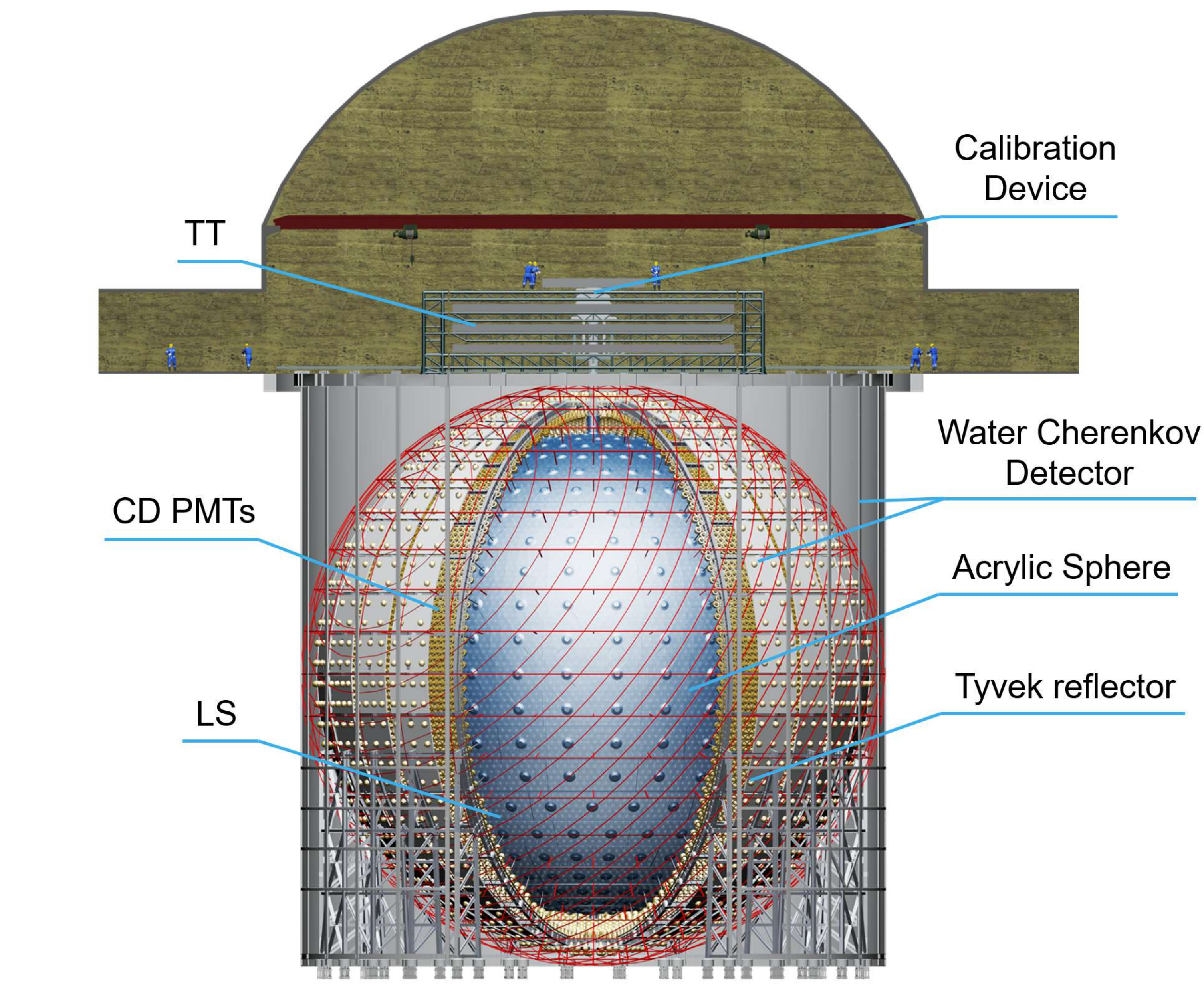}
	\caption{Schematic View of the JUNO Detector.}
	\label{fig:JUNO design}
\end{figure}

The schematic design of JUNO detector is illustrated in Figure~\ref{fig:JUNO design}~\cite{JUNO2}. 
The detector includes the water pool, the CD~\cite{CD}, and the Top Tracker~\cite{TT}. 
The CD is the heart of JUNO experiment and is filled with 20~ktons of liquid scintillator~\cite{zhouxiang_LS1, zhouxiang_LS2} to serve as the target for neutrino detection.
The liquid scintillator is housed within a spherical acrylic vessel, which has a thickness of 120~mm and an inner diameter of 35.4~meters.
This vessel is supported by a spherical stainless steel structure with an inner diameter of 40.1~meters.
To detect photons, the CD is equipped with a total of 17,612 20-inch PMTs and 25,600 3-inch PMTs. 
Surrounding the CD is the water pool containing 35~ktons of highly purified water, which effectively shields the detector from external radioactivity originating from the surrounding rocks. 
The water pool is also instrumental in vetoing cosmic ray muons, with 2,400 20-inch PMTs deployed as part of the water Cherenkov detector.
The Top Tracker, located at the top of the water pool, plays a key role in measuring and vetoing muon tracks~\cite{juno_geo, juno_dynamic}.

\begin{figure}[!htb]
	\includegraphics[width=0.75\hsize]{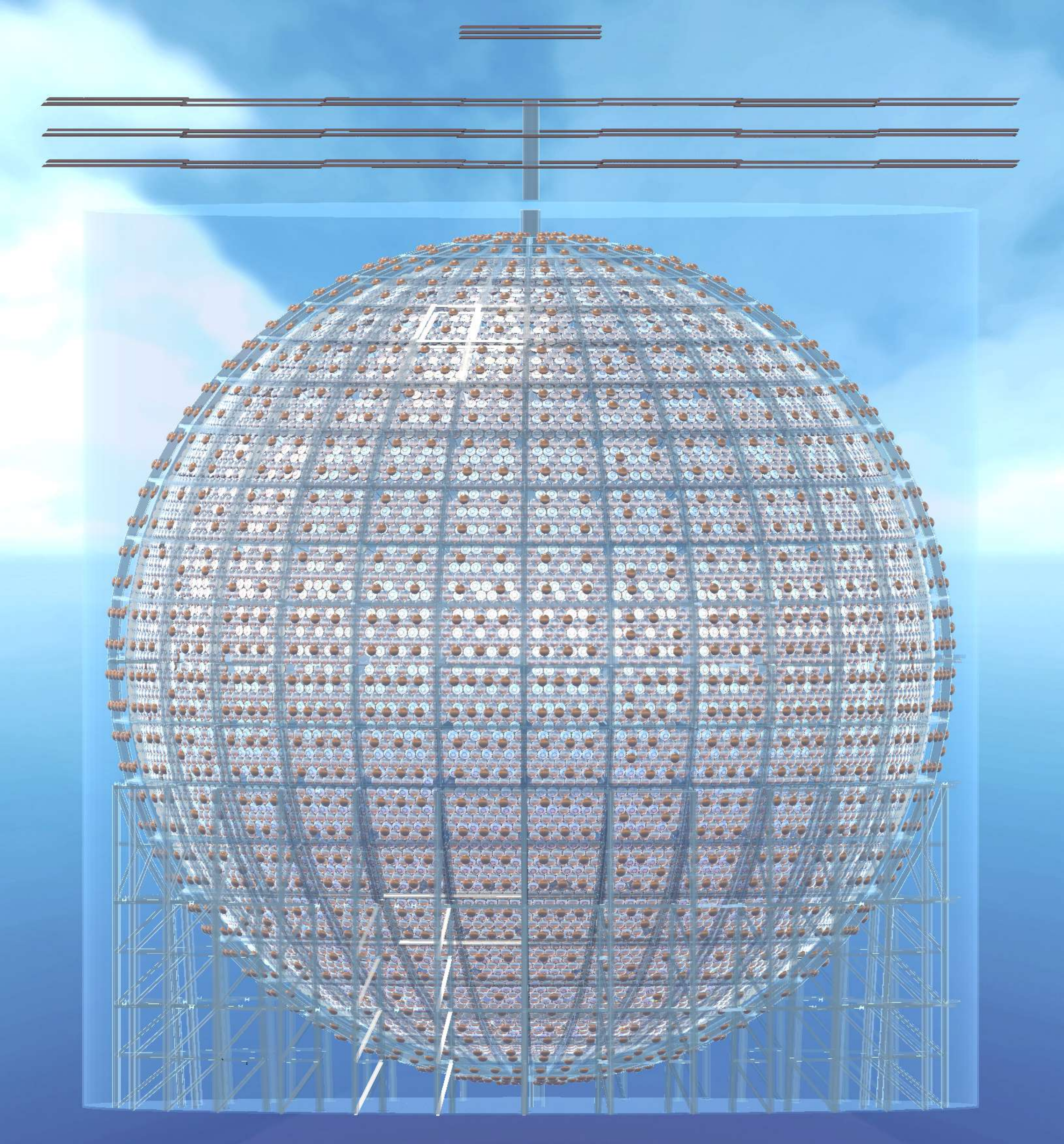}
	\caption{JUNO detector in the VR application.}
	\label{fig:JUNO VR detector in Unity}
\end{figure}

As described in Section~\ref{subsec: Detector Geometry Conversion}, the JUNO detector geometry is converted from the GDML file, and matched between the identifier module and Unity geometry for each detector unit. 
The visualization effects of the whole JUNO detector in VR application is shown in Figure~\ref{fig:JUNO VR detector in Unity}.

The light blue cylindrical structure represents the water pool, with the water pool PMTs positioned outward, as indicated by the yellow portion of the spherical structure.
At the top of the water pool, the reddish-brown structure represents the Top Tracker detector.
From the interior view in the JUNO VR, the spherical acrylic vessel is shown in light gray, as depicted in Figure~\ref{fig:Spatial UI}, although it is close to fully transparent in reality to allow more photons to pass through.
Surrounding this vessel is the stainless steel structure, shown as dark gray in Figure ~\ref{fig:JUNO VR detector in Unity}. 
The CD detector PMTs, oriented toward the center of the sphere, are designed such to receive photons with its photocathodes, so that only the white tail structures of every PMTs are visible in Figure ~\ref{fig:JUNO VR detector in Unity}. 

Owing to the hardware capabilities of Meta Quest~3, there is no requirement to optimize the grid of detector units or replace them with simplified geometric shapes.
Most of the geometric details of the detector units are preserved, achieving effects that are difficult to accomplish in event displays based on ROOT.
Additionally, for the detector units, in order to more closely replicate the effect of real PMTs, we have assigned different material properties to the detector units, including the visualization attributes such as color, reflectivity, and metalicity, to achieve the best display effect.

\subsection{ MC simulation event display }
MC simulation is crucial for detector design and assists physicists in evaluating the detector's performance and tuning reconstruction algorithms.
There are various kinds of signal and backgrounds events in JUNO, while currently we primarily focus on the radioactive backgrounds, IBD signals, and muon events. 

The IBD event, $\nu_e + p \rightarrow e^+ + n$, is the major signal event for detecting electron anti-neutrinos in the JUNO experiment~\cite{JUNO1,JUNO2}.
JUNO identifies and reconstructs IBD events by detecting signals from positron and neutron captures.
This dual-signal characteristic helps effectively identify anti-neutrino signal events while suppressing the huge background events.

\begin{figure}[!htb]
	\includegraphics[width=0.9\hsize]{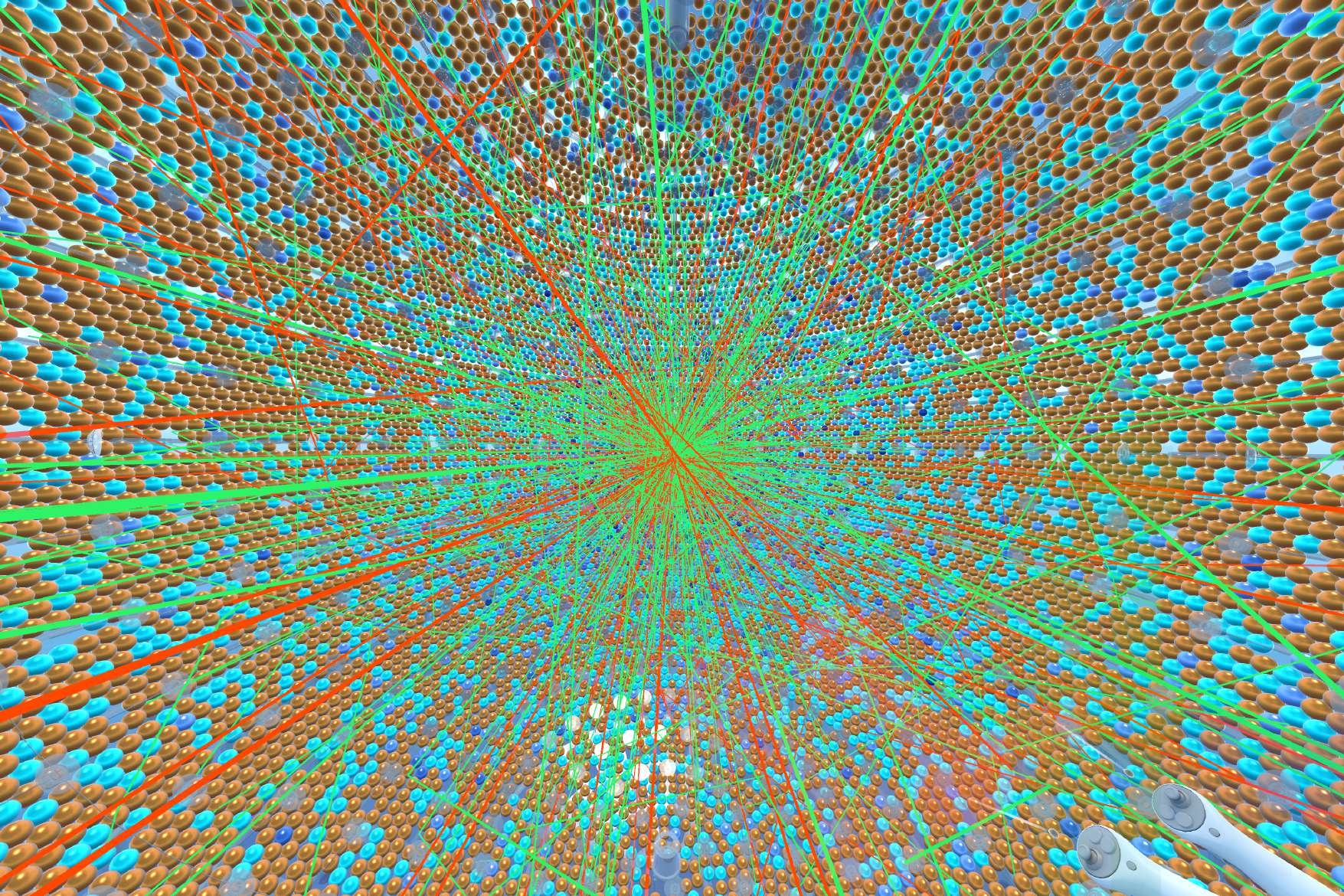}
	\caption{
The event display for a simulated IBD event in JUNO VR application. 
The green lines represent the photon paths of the positron, while the red lines indicate the photon paths of the neutron. 
The yellow spheres represent PMTs that are not triggered, while the spheres with color gradient from light blue to blue indicate the PMTs with an increasing number of hits.
    }
	\label{fig:evt_sim_IBD}
\end{figure}

For the IBD event, there are both positron and neutron signals, whose photon paths are displayed in green and red, respectively, as shown in Figure~\ref{fig:evt_sim_IBD}. 
The detector units that are triggered are color-coded from cyan to dark blue based on the number of hits in the event, with bluer colors indicating a higher number of hits.
The PMTs that are not triggered are displayed in yellow by default. 
Furthermore, in the time evolution of an event, the color of fired PMTs changes with time according to the associated timing information.
The neutron-induced photon paths are delayed by approximately 170 $\mu$s relative to those from the positron, and this delay can be visualized using the time slider in the JUNO VR environment.

\begin{figure}[!htb]
	\includegraphics[width=0.9\hsize]{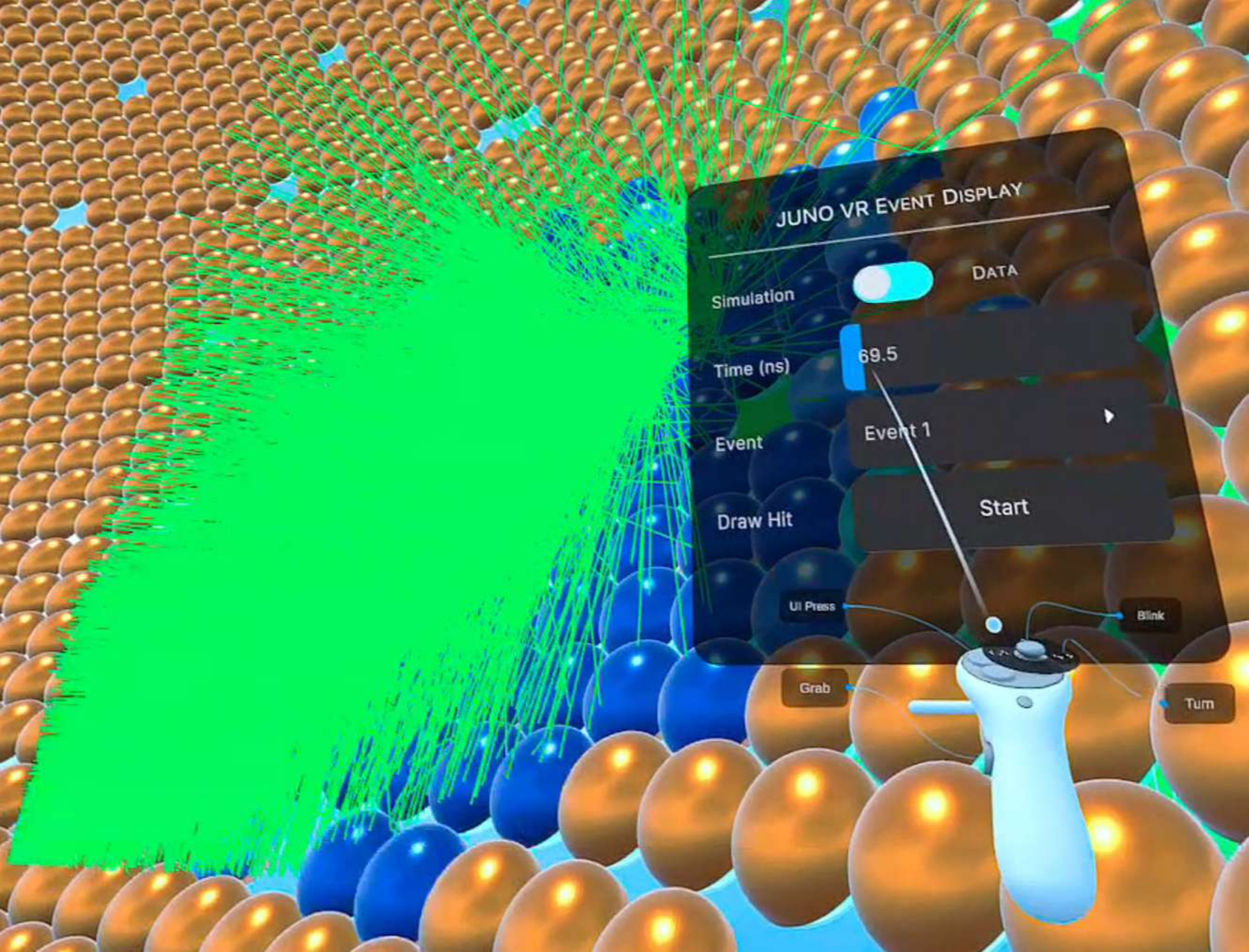}
	\caption{
The event display for a simulated muon event in JUNO VR application.
The green lines represent the photons generated along the path of a muon penetrating the detector.
The controllers and the lasers emitted from the controllers represent the user's interactive control.
    }
	\label{fig:evt_sim_muon}
\end{figure}


One major background event type is the cosmic-ray muon events.
Muons are secondary particles produced by high-energy cosmic rays in the earth's atmosphere, and possess strong penetrating power.
Despite JUNO being located 650~m deep underground, a small fraction of muons can still penetrate the overlying shielding and enter the detector, generating muon events.

Figure~\ref{fig:evt_sim_muon} presents the event information for the simulated muon event.
Photon trajectories are represented by light green lines.
These paths gradually extend over time, depicting the propagation of photons.
In the simulated event shown, the directions of these photon paths may change, indicating their interactions with the detector materials.
For the muon event, as a muon penetrates the detector, it continuously produces photons while depositing its energy in the liquid scintillator.

\begin{figure}[!htb]
	\includegraphics[width=0.9\hsize]{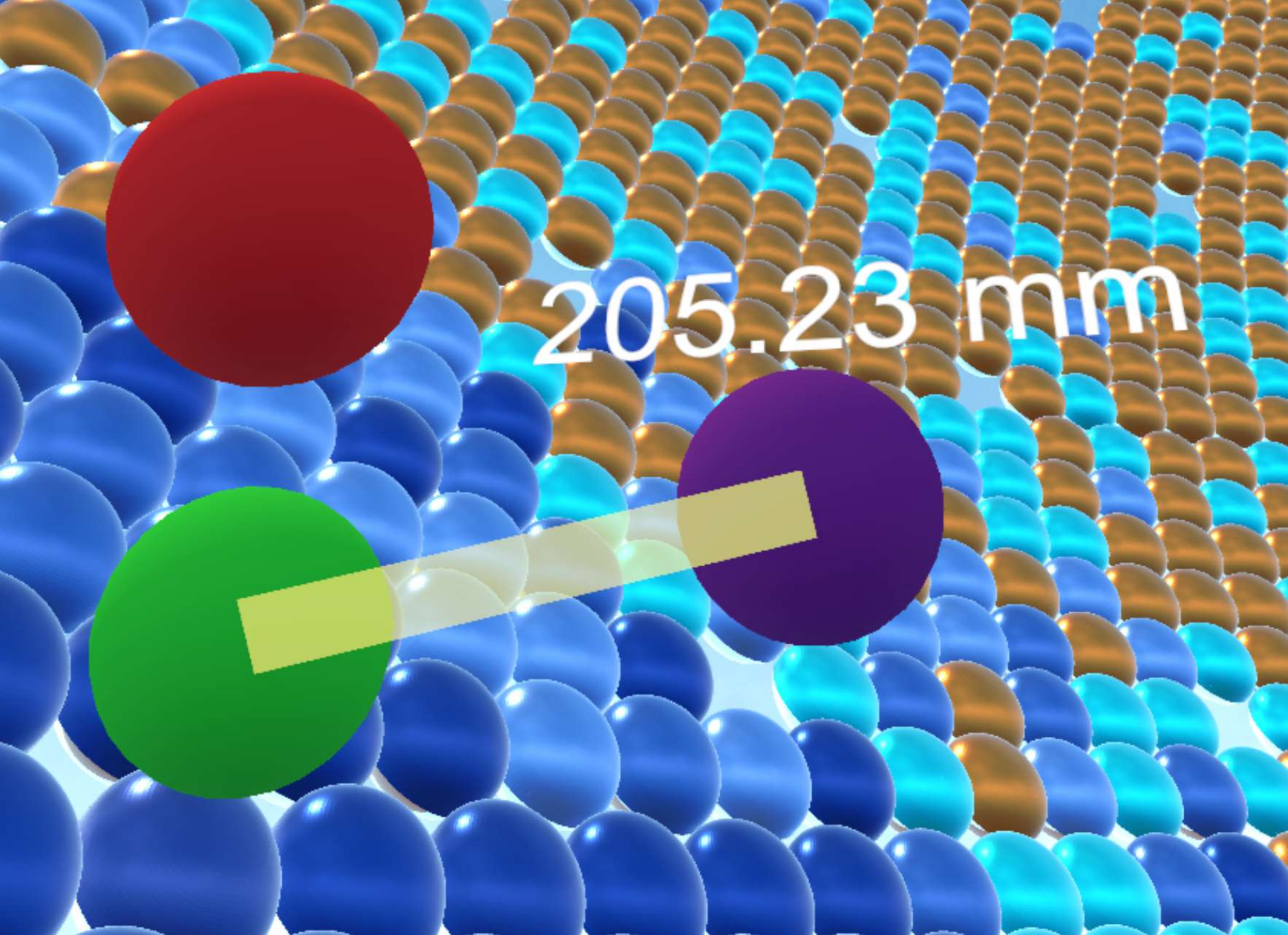}
	\caption{
Comparison of the reconstructed vertex~(purple) with the weighted energy deposit vertex~(green) and the particle production vertex~(red) from MC truth in a simulated event. The yellow line shows the reconstruction bias.
    }
	\label{fig:evt_sim_vertex}
\end{figure}

Event reconstruction plays a key role in JUNO data processing, reconstructing the vertex and energy of an event, which is essential in determining the neutrino mass hierarchy. 
For point-like events like IBD signals, almost all the photon paths originate from the same event vertex. 
Figure~\ref{fig:evt_sim_vertex} shows the reconstructed vertex and the MC truth. 
The initial particle production vertex~(red sphere), derived from MC truth, indicates where the positron is created.
The weighted energy deposit vertex~(green sphere) marks the positron’s annihilation point in the liquid scintillator.
The reconstructed vertex~(purple sphere) is produced by the event reconstruction algorithm.
The reconstruction bias~(light yellow line) represents the discrepancy between the reconstructed vertex and the energy deposit vertex.
A shorter distance indicates a more accurate reconstructed vertex.
In the ideal scenario, the reconstructed vertex will converge to the true vertex.

\subsection{Real data event display}
For the real-data event, we utilize the Data Challenge dataset~\cite{DC1}, whose data structures and processing pipeline are identical to those employed during data taking. This ensures that the software will function seamlessly once the experiment enters formal operation. 
The event composition in this dataset is the same as that in the MC simulation, encompassing radioactive-background events, IBD signals, and muon events.


\begin{figure}[!htb]
	\includegraphics[width=0.9\hsize]{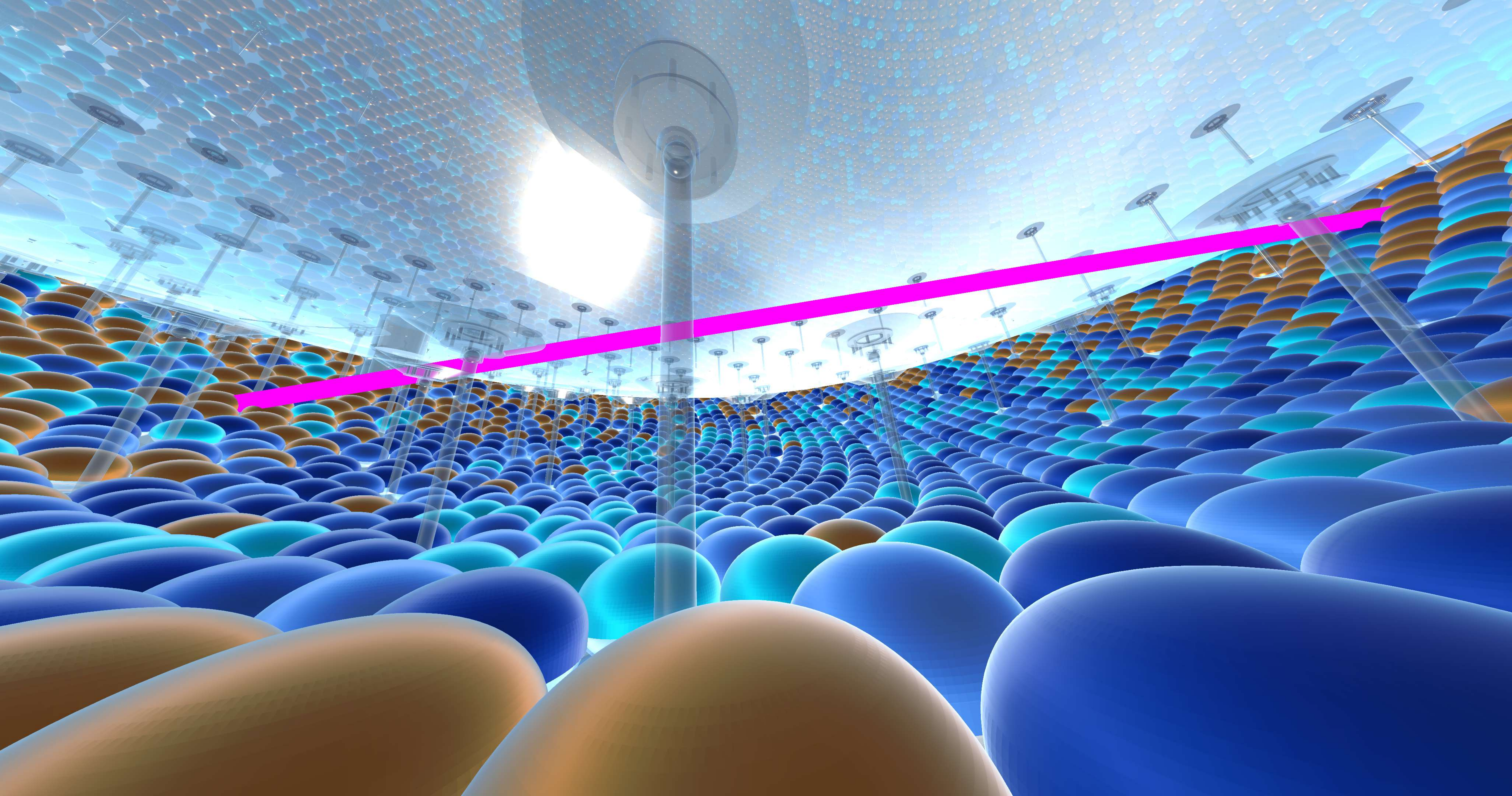}
	\caption{
Display of a reconstructed muon event from datasets in the JUNO VR application.
The translucent part represents the CD acrylic sphere and its supporting components.
The magenta line indicates the reconstructed muon track by connecting the points where the muon enters and exits the JUNO detector.
    }
	\label{evt_DC_muon}
\end{figure}

Figure~\ref{evt_DC_muon} presents event information for a muon event derived from the real data events.
The reconstructed muon travels through the detector along the magenta line.
The left and right sides represent the reconstructed incident and exit points of the muon.
A time offset is established by dividing the track distance by the speed of light.
User can observe the trajectory of muon by the Spatial UI.
Since the exact point of photon emission along the path cannot be determined, photon information is not displayed in this mode. 
Using the reconstructed hit time, the corresponding point on the trajectory is linked to the relevant PMT unit. 
Once the photon particles arrive at the PMT units, those triggered PMTs will change color accordingly.


Moreover, by exploiting Unity’s robust visualization capabilities, a specialized mode is developed to simulate photon paths using particle-like effects instead of simple line trajectories to display the propagation of particles more realistically.

\section{Applications}
\label{sec: Applications}
JUNO VR software provides an immersive interactive experience, allowing user to intuitively understand the detector structure and event information.
Some features and applications of the visualization software are listed below.

\textit{Data quality monitoring.}
The data quality monitoring system~\cite{DQM_ATLAS, DQM_CMS, DQM_BESIII, DQM_dayabay} is designed to identify data issues promptly, ensuring the acquisition of high-quality data.
During the future data-taking phase, event information can be real-time and automatically extracted from the reconstructed files from the data quality monitoring system.
Based on Unity-supported databases, such as SQLite, event information can be transmitted from the data quality monitoring server to JUNO VR software.
This enables immersive visualization of detector operation status and event information during the data-taking phase.
For example, an animation of a real‐time data‐acquisition event is automatically played every 30 seconds.
Through immersive visualization, shifters can easily monitor anomalies, such as hot or dead PMT channels.

\textit{Physics analysis.}
Physical analysis involves in-depth research of neutrino events to extract physical parameters, validate theoretical models, search for rare signals and uncover new phenomena.
This requires detailed analysis of large volumes of complex data.
Through the VR interface, researchers can reconstruct an immersive view of the event in three-dimensional space, allowing them to freely explore the data, observe event details from multiple perspectives, and identify potential patterns and anomalies.

\textit{Outreach.}
For the HEP experiments, their complex theoretical and experimental contents are usually difficult for the public and students to understand.
Based on the VR application, students can understand the structure of JUNO detector and the processing of signal and background events through interactive operations, thereby enhancing engagement and understanding of the physics and principle of the HEP experiments. And the visualization programs including VR stand out in the field of education and public outreach.
Due to Unity's cross-platform support and compatibility with various HMDs, the completed project can be exported to different platforms and utilized with different HMDs, meeting the requirements of various outreach scenarios.

\section{Performance}
\label{sec: Performance}
In experimental evaluations conducted on the mainstream VR device—the Meta Quest~3, the JUNO VR application is capable of processing a variety of event types and demonstrates sufficient computational performance.
During testing, the device’s CPU utilization remains below 70\%, GPU utilization remains below 40\%, and the display maintains a stable refresh rate of 72 frames per second.
The software’s interactive response primarily depends on the event type.
For muon events, which contain a larger volume of hit information, the latency when switching between events is approximately 3 seconds; for IBD and radioactive background events, it is approximately 1 second.

The event display of the JUNO VR application undergoes rigorous testing, and the application is capable of processing both simulated and real data events.

\section{Summary}
The VR technology greatly enhances the visualization effects of HEP experiments.
A JUNO VR application for detector and event visualization is developed using Unity.
By converting GDML to FBX format, efficient construction of the complex detector geometry in Unity is achieved.
An event data conversion interface is created based on matching the detector identifier module and the detector geometry hierarchy in Unity.
Through the Spatial UIs, users can easily control the display of various subsystems for detector and event visualization.

With the ongoing construction of the JUNO experiment, the VR event display software is successfully developed, and more features are expected to be added in the future updates.
VR technology offers an immersive, interactive experience, and it holds great potential in areas such as offline software development, data taking, physics analysis, education and public outreach.

\section*{List of Abbreviations}
\begin{acronym}[DD4hep]  
    \acro{HEP}{High-energy Physics}
    \acro{VR}{Virtual Reality}
    \acro{JUNO}{Jiangmen Underground Neutrino Observatory}
    \acro{TEV}{the Total Event Visualizer}
    \acro{HMDs}{Head-Mounted Displays}
    \acro{Spatial UI}{Spatial User Interface}
    \acro{GDML}{Geometry Description Markup Language}
    \acro{EDM}{Event Data Model}
    \acro{DD4hep}{Detector Description for High-Energy Physics}
    \acro{XML}{Extensible Markup Language}
    \acro{MC}{Monte Carlo}
    \acro{IBD}{Inverse Beta Decay}
    \acro{CD}{Central Detector}
\end{acronym}

\section*{Acknowledgements}
This work was supported by the National Natural Science Foundation of China (Nos. 12175321, W2443004, 11975021, 11675275, and U1932101), Strategic Priority Research Program of the Chinese Academy of Sciences (No. XDA10010900), National Key Research and Development Program of China (Nos. 2023YFA1606000 and 2020YFA0406400), National College Students Science and Technology Innovation Project, and Undergraduate Base Scientific Research Project of Sun Yat-sen University.


\printbibheading[heading=bibintoc]
\printbibliography[heading=none]

\end{document}